\documentclass[elsepaper,11 pt]{article}

 \usepackage[ left=3cm, right=3cm]{geometry}
\usepackage{amsmath}
\usepackage[T1]{fontenc}
\usepackage{amssymb}
\usepackage{mathrsfs} 
\usepackage[english]{babel}
\usepackage[dvips]{graphicx}

\begin{document}

\title{Effect of the fluctuations around mean field for N-body systems with long range interactions}

\author{Y. Chaffi, R. Casta, L. Brenig\\
   \textit{Physique des Systemes Dynamiques, Universite Libre de Bruxelles} }
\date{16 september 2013}

\maketitle

\begin{abstract}
We study the effect of Chandrasekhar and Holstmark's distribution of field fluctuations on the dynamics of N-body systems interacting via Coulomb or Newton gravitational force. We develop an approach based on statistical dynamics first principles whose mathematical framework is similar to the one used by Chandrasekhar and Holstmark for their field fluctuation theory. We use the Picard iteration method to approximate the Hamiltonian dynamics in the short time limit. Neglecting correlations between particles, carrying the thermodynamic limit and assuming that the system is spatially homogeneous, we find a fractional kinetic equation for the velocity distribution. Both, the fractional derivative order and the asymptotic behavior of the solution appear to be directly connected to the $1/r^2$ behavior of the Coulombian or gravitational interaction force over short distances.
\end{abstract}

\textbf{keywords} Holstmark distribution, fractional kinetics, velocity distributions, Levy statistics.

\section{Introduction}

In this article, we intend to merge two different theories concerning N-body systems interacting via Coulomb electrostatic or Newton gravitational forces. The first theory is the kinetic description of such systems based on the Vlasov equation. The second one is the Chandrasekhar-Holtsmark static description of the total field fluctuations. \\
The latter is a static approach of the statistical fluctuations of the total field acting on a given particle interacting with N-1 other identical particles. The interaction corresponds either to the Coulomb or the Newton potential. Holtsmark first developed this theory in the framework of plasma systems \cite{Holtsmark} while Chandrasekhar, inspired by the former, applied it to stellar systems \cite{Chandrasekhar}. In both cases, the position variables of the particles are supposed to be uniformly distributed and uncorrelated. From the probability density of the particle positions the two authors derived the probability density of the total Coulomb or gravitational field acting on one of the particles of the system. In the thermodynamic limit, they rigorously obtained a Lévy stable distribution of index 3/2 for the fluctuations of the total field. This result is nothing else than the application of the generalized Central Limit theorem\cite{GCLT}, the total field being in the above conditions a sum of independant and identically distributed variables.

The second and higher order moments of the total field diverge due to the algebraic long tail of the Lévy $3/2$ distribution. This divergence is essentially caused by the divergence at short distances of the forces in $1/r^2$. As for the average of the total field, it vanishes since the particles have been supposed to be uniformly distributed in position. Though the second and higher order moments diverge, the probability distribution exists and provides informations on the fluctuations of the total field.

These results obtained by Holtsmark and Chandrasekhar are purely static and limited to homogeneous systems. Some extensions to inhomogeneous systems have been derived \cite{kandrup, chavanis}. In these cases the average total field is the so-called Vlasov mean-field and the probability distribution of the total field at a given particle describes the fluctuations around the mean-field.

It is well-known that the action of the mean-field on the time evolution of the one-particle phase-space distribution results in the Vlasov kinetic equation. However, the fluctuations of the total field around the mean-field should also have consequences on that evolution. This is the main object of this article. These fluctuations can be expected to have an effect on the dynamics of the system, at least, in a time regime in which binary collisions still play a negligible role in its evolution. More generally, processes involving Lévy distributions are well known to be connected with the presence of anomalous diffusion and, fractional kinetics \cite{FFP, FKE, LevyFlight}. However,  up to our knowledge, the effect on the dynamics of the Chandrasekhar-Holstmark distribution has not been included in the kinetic description of these systems.\\

In the short time limit, the kinetic description of Coulombian and gravitational systems is based on the Vlasov equation, which describes the reversible evolution of the system due to collective field effects. In this description, the interaction felt by each particle is averaged over the whole system, that is to say, the individual or discrete nature of the particles is smeared out. The importance of the Vlasov equation for these systems stems of course, from the long range nature of both the Coulombian and gravitational interaction. Concerning the numerical simulations, it has been found that starting from arbitrary initial conditions, the system first evolves very quickly, toward the so called quasi-stationary-states(QSS)  which are generally different from the Boltzman distribution. In this first stage of evolution, often referred as violent relaxation \cite{Lynden}, the evolution is assumed to be dominated by collective field effects which are well described by the Vlasov equation. Numerical simulations also shows that these QSS, then, evolve very slowly on a time that scales as $N^{\delta}$, where $N$ is the number of particles and $\delta>1$, until the true thermodynamic equilibrium is reached. It is usually assumed that this last stage of evolution is dominated by binary collisions.\\
	Under this two-stage process scenario, which is the most commonly accepted one \cite{VlasCol1, VlasCol2}, the QSS are naturally interpreted as stationary solutions of the Vlasov equation. However, because of the infinite number of these stationary states, it is not known which one will be selected by the system. Lynden-Bell attempted to solve this issue by introducing a new kind of statistics under which, the most probable QSS is found by entropy maximization  \cite{Lynden}. His theory nonetheless, only works in specific situations \cite{Levin} and, has been revealed inconsistent later on by Lynden-Bell himself and I. Arad due to its non-transitive nature \cite{Lynden2}. Moreover, the same author argues in \cite{Lynden2} that the kinetic description of violent relaxation is probably incomplete, and that a dynamical approach is needed. P.H. Chavannis proposed to add a Fokker-Planck diffusion and friction term to the Vlasov equation in order to describe this transient dynamical stage \cite{chavanis2006quasi}. In this regard, the question we ask is whether there could be any unknown process bridging the two time scales, eventually breaking off the QSS degeneracy, and more particularly, whether that process could be the fluctuations around mean field predicted by Holtsmark and Chandrasekhar.\\

A first attempt at introducing these effects appears in the work of Ebeling et al \cite{ebeling1} in the context of plasma physics.
Out-of-equilibrium probability distribution functions displaying strong non-Gaussian character have indeed been reported \cite{ExpEvAD1,ExpEvAD2,ExpEvAD3}, as well as long range time and/or space correlations in plasma turbulence \cite{ExpEvAD4, ExpEvAD5, ExpEvAD6}, the latter being characteristic of anomalous diffusion. The link with Holtsmark distribution, which is a well known result in plasma physics, has been hypothesized by W. Ebeling et al. In their approach, they write a phenomenological Langevin equation for the velocity with two random forces, a gaussian distributed one representing small angle scattering, and, a Lévy distributed one representing local microfield fluctuations, both processes being assumed to be markovian and stationary. The approach then leads to a fractional Fokker Planck equation, in which the diffusion term has a derivative of a fractional order. Its solution for the velocity distribution is found to be a convoluted Gauss-Lévy, a distribution that has the particularity of having a gaussian core, but behaving asymptotically as a Lévy $3/2$ distribution. Their approach is however not derived from first principles, more specifically, the assumption of stationarity and markovianity of the Lévy process is hard to justify without a rigorous analysis of the microscopic dynamics underlying the process.\\
One of the motivations of the Ebeling and coworkers' approach is the frequently observed discrepancy between the theoretical and experimental nuclear fusion rates in fusion plasmas.ÊThe theoretical probabilities of high velocity particles, that is, the tail of the velocity distribution, appear to be under-estimated. Fusion processes select indeed high-momentum particles that are able to cross the Coulombian barrier, and are thus, extremely sensitive to asymptotic behaviors of velocity distributions. Some work even suggest the use of fusion rates as a probe for tail behaviors of velocity distributions\cite{FusionProbe}.\\

In the present article, we show how a Lévy 3/2 distribution and a fractional order Fokker-Planck equation naturally emerge from a fundamental statistical mechanical description of a charged or gravitational gas. In section 2, we develop a theoretical framework that naturally extends the one used by Chandrasekhar and Holstmark but which includes the microscopic dynamic of particles. In particular, the phase-space one-particle distribution solution of the statistical dynamical problem, $f(\vec{r},\vec{v},t)$, is directly expressed as a functional of initial conditions. In section 3, we propose a very efficient way to approximate the microscopic dynamics in the short time limit based on Picard iterations. In section 4, we expose the result at the first Picard iterates approximation, and explain how the thermodynamic limit can be carried out. The solution displays then clearly, a part involving the free motion of particles, and, one involving the interactions between them. In section 5, we derive an explicit form for the velocity distribution in the case of a spatially homogeneous system. The latter appears to be a convolution of a Levy distribution and of the initial velocity distribution. We moreover derive its equation of evolution, which contains a fractional Laplacian. And finally, in section 6 we will end with some concluding remarks and perspectives for further developments of our approach.

\section{General approach}

Consider a system of $N$ identical interacting particles of mass $m$ of phase-space variables $ \vec{r}_i$, $\vec{v}_i$ ($i=1, \cdots N$). In the remainder of this paper the interacting force will be assumed to be of the following form $\vec{F}(\vec{r}) =\beta \vec{r}/r^3$, where $\beta$ is the coupling constant( $e^2/(4\pi \epsilon_0)$ or $G$ for, respectively, a Coulombian or gravitational force). The probability density of particle $1$ at time $t$ can be formally written as
\begin{equation}
p(\vec{r_1}, \vec{v_1}; t) = \langle \delta (\vec{r_1}-\vec{r_1}(t)) \delta (\vec{v_1}-\vec{v_1}(t)) \rangle \label{eqchaf_1}
\end{equation}
where $\langle \cdot \rangle $ denotes the averaging process over the initial conditions for the N-body system which are random variables distributed according to the $N$ particle probability density joint distribution function at time $0$, $P_N( \Gamma ; 0)$ ($\Gamma$ is the collection of initial positions and velocities \{$\vec{x_i}, \vec{u_i}\}$). The functions $\vec{v_1}(t)$  and $\vec{r_1}(t)$ stand for the dynamical variables, solutions of Hamilton's equations,

\begin{align}
\frac{d \vec{r}_i(t) }{dt}\,  & =  \vec{v}_i(t)   \label{hamiltonian}  \\
\frac{d \vec{v}_i(t) }{dt} & =  \frac{1}{m}  \sum_{j=1, j\neq i}^N \vec{F}\left(\vec{r}_i(t)-\vec{r}_j(t)\right)\,\,\,  . & \notag
\end{align}
Their solutions for particle $1$ can be obtained from the integral equation
\begin{align}
\vec{v_1}(t) & =  \vec{u} _1+ \frac{1}{m} \int_0^t  \sum_{j=2}^N \vec{F}\left(\vec{r}_1(t')-\vec{r}_j(t')\right) dt'  \label{eqchaf_2} & \\
\vec{r_1}(t) & = \vec{x}_1+ \frac{1}{m} \int_0^t \vec{v}_1(t')  dt'  . & \notag
\end{align}
It is important to note that the initial positions and velocities of all particles being random variables, $\vec{r}_1(t) $ and $ \vec{v}_1(t) $ being functions of them are, hence, also random variables. Now writing down explicitly the averaging process and using the following identity $\delta(\vec{x}) =  (2\pi)^{-3} \int  d^3\xi \,\, e^{i\vec{\xi} \cdot \vec{x}}$, one obtains the following expression from equation  \eqref{eqchaf_1},

\begin{align}
p(\vec{r_1}, \vec{v_1}; t) & =\int \frac{ d^3 \xi d^3 \zeta}{(2\pi)^6} e^{i\vec{\xi}\cdot \vec{r}_1+ i\vec{\zeta}\cdot \vec{v}_1  }  \int d^{6N}  \Gamma \,\,\,  P_N(\Gamma,0) \,\,\,    e^{-i \vec{\xi} \cdot  \vec{x}_1 - i \vec{\zeta} \cdot \vec{u}_1 }  \label{eq4} & \\
&  \times \exp{ \left( i \vec{\xi} \cdot \frac{1}{m}  \int_0^t \vec{v}_1(t')  dt'  \right)}   \prod_{j=2}^N \exp\left( i\vec{\zeta}\cdot \frac{1}{m} \int_0^t dt'  \vec{F}(\vec{r}_1(t')-\vec{r}_j(t')) \right)  &  \notag
\end{align}
where $ d^{6N} \Gamma =   d^{3}x_1 d^{3}u_1  \cdots  d^{3}x_N d^{3}u_N $ is the volume element of the averaging integral over all the possible initial points, ie., over the whole N-body phase-space. \\
Equation \eqref{eq4}, despite its formal character, provides the full and exact solution of the nonequilibrium statistical dynamical problem and is thus equivalent to the solution of the Liouville equation integrated over all phase-space variables except particle $1$. Indeed, as we discuss later, the probability density $p(\vec{r}_1, \vec{v}_1; t)$ is, up to a factor N, equivalent to the reduced one-particle distribution function. The essential difference of this approach with standard theories of statistical dynamics is that the statistical character is introduced only through the initial conditions. This feature allows us to incorporate the particle dynamics in its integrated form. \\
This approach has several interesting features. First, it shows precisely where the solutions of the microscopic dynamics enter in the description. Second, as will be shown in the next section, it allows us to perform different approximations and perturbative schemes which would not be straightforward using the standard Bogolyubov-Born-Green-Kirkwood-Yvon(BBGKY) approach.
Before closing this section, let us mention that this method can of course be used to express the $s$ particle probability density for any $s=1,2 \cdots N$,
\begin{equation}
 p_s(\vec{r_1}, \vec{v_1}, \cdots \vec{r_s}, \vec{v_s} ;  t)= \langle \,\,\, \prod_{j=1}^s \delta (\vec{r_j}-\vec{r_j}(t)) \delta (\vec{v_j}-\vec{v_j}(t)) \,\,\, \rangle
\end{equation}
in terms of the solutions of the Hamilton's equations, in complete analogy with equation \eqref{eq4}. \\

\section{Approximation schemes}

In order to make formula \eqref{eq4} explicit, one has to specify the initial statistical state and to introduce an explicit solution of the equations of motions for the positions and velocities of the N particles. In this work, we will consider that the system is initially in an uncorrelated state\footnote{Without loss of generalities, one could also consider the effect of initial \emph{s-particles} correlations ($s=2,3 \cdots N$) to the evolution of the probability distribution function $p(\vec{r_1}, \vec{v_1}; t)$.}, leading to the factorization of the initial probability distribution function,

\begin{equation}
P_N(\vec{x}_1,\vec{u}_1,\vec{x}_2,\vec{u}_2 \cdots \vec{x}_N,\vec{u}_N ; 0)= \prod_{i=1}^N p(\vec{x}_i,\vec{u}_i; 0) . \label{factored}
\end{equation}
Of course, this initial factorization is by no means necessary and we will report in a subsequent article on how our approach generalizes  when correlations are present in the initial statistics.\\
Concerning the solutions of the dynamical equations, we use a very efficient method called Picard iterations. This method provides successive approximations or iterates of the solution of a system of ordinary differential equations(ODE), in which the iterate $n+1$ is given as a functional of the $n-th$ iterate. The given set of iterates will converge to the exact solution if the generator of the evolution satisfies the Lipshitz condition,
\begin{equation}
\| \vec{F}(\vec{\Gamma}_1) - \vec{F}(\vec{\Gamma}_2 ) \|< L \| \vec{\Gamma}_1 - \vec{\Gamma}_2  \| \label{lipschitz}
\end{equation}
where the generator $\vec{F}$ is the right hand side of the Hamiltonian equations \eqref{hamiltonian}, $\vec{\Gamma}_1, \vec{\Gamma}_2$ are arbitrary phase-space points and $L$ is the Lipschitz constant.\\ 
At this point, a remark needs to be made regarding the existence of a Lipschitz constant $L$. For $1/r^2$ forces or, for that matter, any force that diverges for arbitrarily short distances, the breakdown of condition \eqref{lipschitz} is indeed expected whenever two given particles of the system get arbitrarily close during the evolution. However, when dealing with statistical dynamics one needs only to consider the likelihood of such an event. It has been shown in \cite{Saari} that the subset of all initial conditions leading to it in a finite time is a zero measure set and thus, is not significant in a statistical description. So, if one considers the evolution of the system on short times, one can assume the existence of a Lipschitz constant.\\
The Picard iteration method has been so far almost exclusively used as a theoretical tool to investigate uniqueness and existence properties of system of ordinary differential equations. More recently it has been used successfully for gravitational N-body simulations \cite{Pruett}. It proves in our case to be a very effective and analytically tractable tool to approximate the formal solutions \eqref{eqchaf_2}. The expression of the $n+1$ iterate of the dynamics is

\begin{align}
& \vec{v}_i^{(n+1)}(t)= \vec{u}_i + \frac{1}{m} \int_0^t \sum_{j\neq i}^N \vec{F}\left(\vec{r}_i^{(n)}(t')-\vec{r}_j^{(n)}(t')\right) dt'  & \\
& \vec{r}_i^{(n+1)}(t)= \vec{x}_i + \int_0^t  \vec{v}_i^{(n)}(t') dt' \,\,. & \notag
\end{align}
\begin{align}
  \mbox{with} & \,\, \vec{v_i}^{(0)}(t) = \vec{u_i} & \\
 &  \vec{r_i}^{(0)}(t) = \vec{x_i} & . \notag
\end{align}
With this scheme we can, thus, get successive approximations (or iterates) of the distribution function \eqref{eqchaf_1} which, for instance, at order $k$ leads to
\begin{equation}
p^{(k)}(\vec{r_1}, \vec{v_1}; t) =  \langle \delta (\vec{r_1}-\vec{r_1}^{(k)}(t)) \delta (\vec{v_1}-\vec{v_1}^{(k)}(t)) \rangle .
\end{equation}
It is important to note, however, that Picard iterations are valid only over short time intervals, which, as expected, increase as one proceeds with further iterates. The characteristic time $t_c$ under which a given iterate will be valid depends on the Lipshitz constant $L$ which itself, defines the inverse of the square of a time. In the case of the first iterate for instance, $L$ defines an upper bound on the error, so given it being sufficiently small or, equivalently considering sufficiently short times will keep the second iterate correction negligible, so one needs not consider it. Thus, beyond the existence of a Lipschitz constant discussed above, the choice of a physically relevant one appears indeed to be crucial for estimating the time validity of our approach. This constant is generally evaluated as the supremum among the components of the tensor gradient of the generator of the system of ordinary differential equations \cite{ODE}. The so called generator being of course, in our case, the RHS of Hamilton's equations \eqref{hamiltonian}.\\

Evaluating the supremum of the aforementioned gradient, which, among its different components, involves the spatial derivatives of the force, - others being either constants or zero -  requires the introduction of a minimum distance between particles. The latter is provided, once again by using a statistical argument, by $n^{-1/3}$, where $n=N/V$ is the average number density of particles in the system. One then finds in the case of a Coulombian or gravitational force that $L=((\beta n)/m)$, which corresponds to the square of either, the plasma frequency or the inverse of Jean's time, respectively. The time validity of the first iterate is thus given by $t_c<<\tau$ where $\tau = L ^{-1/2}= ((\beta n)/m)^{-1/2}$.\\

We expose below our results for the first iterate.

\section{The first order approximation}

We now show how, together, the hypothesis of an initial uncorrelated state \eqref{factored} and the first Picard iterates approximation of the dynamics, lead to a considerable simplification of the expression \eqref{eq4}. The first Picard iteration of the solution to equations \eqref{hamiltonian} is
\begin{align}
& \vec{v_1}^{(1)}(t)= \vec{u}_1 +   \sum_{j=2}^N \vec{\Delta}_j  & \label{firstiterate} \\
& \vec{r_1}^{(1)}(t)= \vec{x}_1 + \vec{u}_1 t \notag  &
\end{align}
where $\vec{\Delta}_j= \vec{F}(\vec{x}_1-\vec{x}_j)t/m $ is the short time approximation of the velocity deflection of particle $1$ due to its interaction with particle $j$. Due to its dependence on the random initial position, $\vec{\Delta}_j$ is of course, also a random variable. One then readily obtains for the probability density distribution,
\begin{align}
p^{(1)}(\vec{r_1}, \vec{v_1}; t) & =\int \frac{ d^3 \xi d^3 \zeta}{(2\pi)^6} e^{i\vec{\xi}\cdot \vec{r}_1+ i\vec{\zeta}\cdot \vec{v}_1  }  \int d^3 x_1 d^3 u_1  \,\,\,   p(\vec{x}_1,\vec{u}_1; 0)   e^{-i \vec{\xi} \cdot  (\vec{x}_1+ \vec{u}_1 t) - i \vec{\zeta} \cdot \vec{u}_1 } \label{eq1} &  \\
&  \times \prod_{j=2}^{N} \tilde{W}_{\vec{\zeta}}(\vec{\Delta}_j ; \vec{x}_1 , t ) \,\,\, &  \notag
\end{align}
where
\begin{equation}
\tilde{W}_{\vec{\zeta}}(\vec{\Delta}_j ; \vec{x}_1 , t ) =  \int d^3 x_j d^3 u_j \,\,\,  p(\vec{x}_j,\vec{u}_j; 0)  \exp(  i\vec{\zeta}\cdot \vec{\Delta}_j )
\end{equation}
 is the characteristic function of the random $\vec{\Delta}_j$. One then clearly sees that as a consequence of the above hypotheses, all $\vec{\Delta}_j$ become independent identically distributed random variables.\\
 Using the normalization condition of $p(\vec{x},\vec{u}; 0)$, in the same way Chandrasekhar followed to get the Holtsmark distribution in \cite{Chandrasekhar}, the product of these characteristic functions may be written in the following form,
 \begin{equation}
 \prod_{j=2}^{N} \tilde{W}_{\vec{\zeta}}(\vec{\Delta}_j ; \vec{x}_1 , t )= \left\{ 1 + \int d^3 x d^3 u \,\,\,  p(\vec{x},\vec{u}; 0) \left( \exp(  i\vec{\zeta}\cdot \vec{\Delta} )-1\right) \right\}^{N-1}
 \end{equation}
 Now, introducing the one-particle phase-space distribution function $f(\vec{r}, \vec{v}) = N p(\vec{r}, \vec{v})$ which is normalized to $N$, it appears clearly that only the interaction term above involves the number of particle as the two $1/N$ factors cancel each other out in the first line of \eqref{eq1}. Performing the thermodynamic limit then leads to
 \begin{equation}
 \lim_{N \to \infty} \prod_{j=2}^{N} \tilde{W}_{\vec{\zeta}}(\vec{\Delta}_j ; \vec{x}_1 , t ) = \exp  \left(  \int d^3 x d^3 u \,\,\,  f(\vec{x},\vec{u}; 0) \left[ \exp(  i\vec{\zeta}\cdot \vec{\Delta} )-1\right] \right),
 \end{equation}
 where we used
 \begin{equation}
\lim_{N \to \infty} \left(1 + x/N\right)^N= \exp(x).
 \end{equation}
 Expression \eqref{eq1} then becomes,
\begin{align}
 f^{(1)}(\vec{r_1}, \vec{v_1}, t) &  = \int\frac{ d^3 \xi d^3 \zeta}{(2\pi)^6} e^{i\vec{\xi}\cdot \vec{r}_1+ i\vec{\zeta}\cdot \vec{v}_1  }\,\,\,\, \int d^3 x_1d^3u_1\,\,\,\, f(\vec{x_1} , \vec{u_1},0 ) e^{-i \vec{\xi} \cdot  (\vec{x}_1+ \vec{u}_1 t)- i \vec{\zeta} \cdot \vec{u}_1 }  & \\
&   \times \,\,\,\, \exp \left( \int d^3 x \,\, n(\vec{x},0)\left[ e^{-i\vec{\zeta}\cdot \frac{t}{m} \vec{F}(\vec{x}_1 -\vec{x})} -1 \right] \right) \,\,  \notag &
\end{align}
and where the local number density is given by $ n(\vec{x},0)  =  \int  d^3 u \,\,\,  f(\vec{x},\vec{u}; 0)$. \\
Moreover, taking into account that $\int d^3 \xi \,\, \exp \left( i\vec{\xi}\cdot (\vec{r}_1 -\vec{x}_1- \vec{u}_1 t ) \right) = (2\pi)^3 \delta (\vec{r}_1 -\vec{x}_1- \vec{u}_1 t) $ one can then perform the integral over $\vec{x_1}$,
\begin{align}
 f^{(1)}(\vec{r}, \vec{v}, t) &  = \int \frac{d^3 \zeta}{(2\pi)^3} \,\,\,\, e^{ i\vec{\zeta}\cdot \vec{v}  }\,\,\,\, \int d^3u\,\,\,\, f(\vec{r} -\vec{u}t , \vec{u},0 ) e^{-i \vec{\zeta} \cdot \vec{u}} \label{yass1}  &  \\
&   \times \,\,\,\, \exp \left( \int d^3 x \,\, n(\vec{x},0)\left[ e^{-i\vec{\zeta}\cdot \frac{t}{m} \vec{F}(\vec{r} -\vec{u}t-\vec{x})} -1 \right] \right) \,\,  \notag &
\end{align}
where we dropped the subscript $1$.\\
This expression has a clear physical meaning, the term in the RHS of the first line corresponds to the Fourier transform of a solution for a pure ballistic motion. On the other hand, the term in the second line represents the effect of the interactions. Formally speaking it is the Fourier transform of the distribution of the sum of velocity deflections felt by particle $1$ due to the interactions with the $N-1$ other particles, conditioned on it being located at the phase-space point $(\vec{r} -\vec{u}t,\vec{u} )$. Physically, this latter quantity does not describe a localized collision process, it indeed only depends on the initial particle density $n(\vec{x},0)$ and does not depend on the relative velocity of the interacting partners, it rather describes a collective process that takes into account the full range of the interaction.\\
It is noteworthy that in the above derivation leading to equation \eqref{yass1}, the thermodynamic limit can be carried out on exactly without specifying the range of the force. \\
We now present a more explicit form of equation  \eqref{yass1} that can be obtained if one restricts our scope to spatially homogeneous systems.

\section{The homogeneous approximation}

In this approximation the one-particle distribution function reduces to $f(\vec{r}, \vec{v};t)= n \varphi (\vec{v};t)$ and the particle density becomes a constant $n(\vec{x})=n=N/V$. The expression for the velocity distribution then simplifies to:

\begin{align}
 \varphi^{(1)}( \vec{v}, t) &  = \int \frac{d^3 \zeta}{(2\pi)^3} \,\,\,\, e^{ i\vec{\zeta}\cdot \vec{v}  }\,\,\,\, \int d^3u\,\,\,\, \varphi( \vec{u},0 ) e^{-i \vec{\zeta} \cdot \vec{u}}  & \\
&   \times \,\,\,\, \exp \left( -n \int d^3 r \,\, \left[ e^{-i\vec{\zeta}\cdot \frac{t}{m} \vec{F}(\vec{r} )} -1 \right] \right) \,\,  \notag &
\end{align}
with the change of variable $\vec{r}=\vec{r}_1 -\vec{u}t-\vec{x}$.
We now assume a Coulombian or gravitational force of the functional form $\vec{F}(\vec{r})=\beta \vec{r}/r^3$ with the coupling constant $\beta$. The integral remaining in the second line is then strictly identical to that appearing in Chandrasekhar's derivation of the Holtsmark distribution \cite{Chandrasekhar}. One then finds
\begin{equation}
 \varphi^{(1)}( \vec{v}, t)   = \int \frac{d^3 \zeta}{(2\pi)^3} \,\,\,\, e^{ i\vec{\zeta}\cdot \vec{v}  }\,\,\,\, \int d^3u\,\,\,\, \varphi( \vec{u},0 )\,\, e^{-i \vec{\zeta} \cdot \vec{u}} \,\,  e^{ - (c(t) \zeta)^{3/2} } \,\, \label{result}
\end{equation}
or in the rather more explicit form,
\begin{equation}
 \varphi^{(1)}(\vec{v}, t)   =  \int d^3u\,\,\,\, L_{3/2}\left( c(t), \vec{v}-\vec{u} \right) \varphi(\vec{u}, 0) \label{homoresult}
\end{equation}
where $c(t)=(\frac{4 n}{15})^{\frac{2}{3}}\,\, \frac{2 \pi \beta t}{m}$ and $L_{3/2}( c, x )$ is the Lévy distribution of index $3/2$ and parameter $c$ of the random variable $x$. \\
From equation \eqref{homoresult}, we see that, for a spatially homogeneous system, the velocity distribution is a convolution of the initial velocity distribution with a Lévy distribution. In probabilistic terms, the random velocity $\vec{v}$ at time $t$ is the sum of two random variables, namely the initial velocity $\vec{u}$ and the total velocity deflexion $\vec{\Delta}_{tot}$ due to the interaction with all the other particles. These two become independent as a consequence of the spatial translational invariance. In the opposite case, it would be expected indeed that the interaction felt by the particle would depend on both its initial velocity and also on its initial position, in which case, the explicit form \eqref{homoresult} as a convolution of two distribution functions could not be obtained.\\
The velocity distribution given by equation \eqref{homoresult} can be shown to have an algebraic long tail in $v^{-5/2}$. This particular feature, which is common to both plasmas and gravitational gases, is a direct consequence of the $!/r^2$ divergence of the interaction force at short distances. A similar result is expected for any interaction force having that kind of divergence at the origin.. \\
An equation of evolution can be derived in a straightforward manner by taking the time derivative of \eqref{result},
\begin{equation}
\partial_t \varphi^{(1)}(\vec{v}, t) = - D(t)   (-\triangle_{\vec{v}})^{3/4} \,\, \varphi^{(1)}(\vec{v}, t) \label{FrakFokk}
\end{equation}
where $D(t)=\frac{3}{2} \frac{(c(t))^{3/2}}{t}$ and the following definition of the fractional power of the Laplacian operator \cite{fractional} has been used,
\begin{equation}
\zeta^{3/2} e^{i\vec{\zeta}\cdot \vec{v}  } = (-\triangle_{\vec{v}}) ^{3/4} \,\,   e^{i\vec{\zeta}\cdot \vec{v}  } \,\, .
\end{equation}

 Equation \eqref{FrakFokk} is a fractional Fokker-Planck equation without drift term. It describes an anomalous diffusion process in velocity space. The presence of this fractional term is, as expected, directly related to the algebraic decay of the solution \eqref{homoresult} and is thus, also a consequence of the behaviour of the force over short distances.\\
 Another noteworthy feature of equation \eqref{FrakFokk}, is the proportionality of $D(t)$ - which plays here the role of a diffusion coefficient - in the coupling constant $\beta$ to the power $3/2$. Morever, it is also proportional to the particle number density $n$, which suggest that the process is essentially driven by nearby particles, similarly to the static case, where, the largest contribution to Holtsmark distribution comes indeed from close particles. All these features converge to the fact that the fractional power of the Laplacian appearing in equation (22) is a non-perturbative result stemming from the short range divergence of the Coulomb and gravitational forces. 

\section{Results and discussion}

The combined use of formula \eqref{eqchaf_1} for the one-particle phase-space probability density, the integral form of the equations of motion \eqref{eqchaf_2} and the Fourier representation of the Dirac delta led us to expression \eqref{eq4}. This equation is the starting point of our approach. It can be made explicit by inserting in it different approximations of the solution of the equations of motion.

The main result of this article, that the velocity distribution at time $t$ is a convolution given by equation \eqref{homoresult} of the initial velocity distribution with a Lévy-3/2 stable distribution, is obtained assuming:
\begin{enumerate}
\item An uncorrelated and spatially homogeneous N-particle distribution at time zero
\item The first order Picard iteration solution of the equation of motion
\item The thermodynamic limit $N \rightarrow \infty$, $V\rightarrow \infty$, $N/V=n$ constant.
\end{enumerate}

We found also that this velocity distribution obeys a closed fractional diffusion equation, equation \eqref{FrakFokk}, in the velocity space. The fractional Laplacian term appearing in its right-hand-side represents the effect of the fluctuations of the total field. These fluctuations are defined around the average total field acting on the particle, i.e. the Vlasov mean-field. However, due to the spatial homogeneity of the system studied here, the mean-field field is exactly zero. In the Vlasov approximation, the velocity distribution is, thus, time invariant for a homogeneous system. In contrast, by taking account of the fluctuations of the total field we obtain a non-trivial time-evolution for that distribution : a fractional diffusive behaviour. This evolution is irreversible and tends toward an asymptotically flat distribution. A Lyapunov function can even be defined for the evolution equation  \eqref{FrakFokk} for which an H-theorem can be established. We do not report this result here for the sake of brevity.

As we already noticed, another important feature should be stressed here. The fractional diffusion term in  \eqref{FrakFokk} is proportional to the power 3/2 of the coupling constant $\beta$. This non-analytical dependance in the coupling constant reflects a non-perturbative property of our derivation. Indeed, the fractional Laplacian or the equivalent fact that a Lévy-3/2 distribution appears in our derivation stems from the important probability of large values of the fluctuations of the total field. These are responsible for the fat tail of the Lévy-3/2 velocity distribution. In turn, these large values come from the force field exerted by the particles that are very near the test-particle. The divergence of the Coulomb or Newton potential at short distances is, thus, the main cause of the appearance of the fractional diffusion term in  \eqref{FrakFokk}. One would recover the same term for any potential that behaves like 1/r at short distances.

One should remark that such a non-perturbative contribution would be very difficult to discover from an analysis based on the BBGKY representation of statistical mechanics. Indeed, truncations of that hierarchy of equations for the reduced phase-space distributions of increasing orders are equivalent to perturbative approximations for the evolution equation of the one-particle distribution.

Before ending this chapter, we should stress again that the above results equations (18), (20) and subsequent are only valid for short times. This comes from the fact that we used the first Picard iteration of the solution to the equations of motion. The validity time is related to the inverse of the Lipschitz constant and is of the order of the inverse of the plasma frequency for plamas or of the Jeans instability constant for gravitational gases. This time is much shorter than the relaxation time that spans the collisional evolution of the systems toward thermal equilibrium.

\section{Conclusions and perspectives}

We propose in this work a new approach of statistical mechanics that allows for a certain kind of non-perturbative treatment of  N-body systems governed by diverging forces at short distances. This treatment permits to take into account the effect of the large fluctuations of the total force field on the time evolution of the systems.

The results we present here are obtained for systems satisfying very strong conditions such as no velocity and position correlation between the particles at time zero, spatial homogeneity and short evolution times. However, these are preliminary results and we hope to be able to extend this theory to initial conditions with correlations and to spatially inhomogeneous systems.

Moreover, that theory should lead to similar results for systems governed by other types of interaction with divergent behaviour at short distances.

This approach should also be confronted to the challenge of describing the collisional evolution toward thermal equilibrium, that is, the long time evolution of the system. This implies that both the non-perturbative contributions describing the effects of nearby particles and the perturbative ones related to far away particles should simultaneously be taken into account in a systematic way.\\

\textbf{Acknowledgements} \\
The authors are grateful to Drs. T.M. Rocha Filho, A.Figueiredo, Y.Elskens, D. Escandes, for fruitful and enlightning discussions during this work.

\bibliographystyle{amsplain}
\bibliography{bibliography}

\end{document}